\documentclass[pra,preprint,amssymb]{revtex4}
\usepackage{graphicx}
\usepackage{bm}
\usepackage{amsmath,amssymb,fancybox}
\usepackage{color}

\begin{document}
\title{Quantum state and circuit distinguishability
with single-qubit measurements}
\author{Tomoyuki Morimae}
\email{morimae@gunma-u.ac.jp}
\affiliation{ASRLD Unit, Gunma University, 1-5-1 Tenjin-cho Kiryu-shi
Gunma-ken, 376-0052, Japan}

\begin{abstract}
We show that the Quantum State Distinguishability (QSD),
which is a QSZK-complete problem,
and the Quantum Circuit Distinguishability (QCD),
which is a QIP-complete problem,
can be solved by the verifier
who can perform only single-qubit measurements.
To show these results, we use  
measurement-based quantum computing:
the honest prover sends a graph state to
the verifier, and the verifier can perform universal
quantum computing on it with only single-qubit measurements.
If the prover is malicious, he does not necessarily generate the correct
graph state, but the verifier can verify the correctness of the
graph state by measuring the stabilizer operators.
\end{abstract}
\pacs{03.67.-a}
\maketitle  

\section{Introduction}
Measurement-based quantum computing~\cite{MBQC} is a 
new model of quantum computing
where universal quantum computing can be realized with
only adaptive single-qubit measurements on a certain
entangled state such as the graph state.
Several applications of
measurement-based quantum computing in quantum computational
complexity theory have been obtained recently.
For example, Ref.~\cite{Matt} used measurement-based quantum
computing to construct a multiprover
interactive proof system for BQP with a classical verifier.
Furthermore, Refs.~\cite{MNS,QAMsingle} used measurement-based quantum
computing to show that the verifier needs only single-qubit measurements
in QMA and QAM.
The basic idea in these results is the verification of the
graph state: prover(s) generate the graph state,
and the verifier performs measurement-based quantum computing on it.
By checking the stabilizer operators, the verifier
can verify the correctness of the graph state.
The idea of testing stabilizer operators was also used in Refs.~\cite{FV,Ji}
to construct multiprover interactive proof systems for local
Hamiltonian problems.

In this paper, we consider
two promise problems,
Quantum State Distinguishability (QSD)~\cite{WatrousQSZK},
which is QSZK-complete, and
Quantum Circuit Distinguishability (QCD)~\cite{Rosgen},
which is QIP-complete.
By using the idea of testing stabilizer operators,
we show that these problems can be solved by the verifier
who can do only single-qubit measurements.
Proofs are similar to those of Refs.~\cite{MNS,QAMsingle} for QMA and QAM,
but several new considerations are required since in protocols
to solve QSD and QCD some parts of graph states are kept by
the prover.

\subsection{QSD}
{\bf Definition}:
Quantum State Distinguishability 
(QSD$_{\alpha,\beta}$)~\cite{WatrousQSZK}.
\begin{itemize}
\item
Input:
Quantum circuits $Q_0$ and $Q_1$ each acting on $m$ qubits
and having $k$ specified output qubits.
\item
Promise:
Let $\rho_a$ ($a\in\{0,1\}$) be the mixed state obtained by 
tracing out the non-output qubits of
$Q_a|0^m\rangle$.
We have either
$\frac{1}{2}\|\rho_0-\rho_1\|_1\ge \beta$ or
$\frac{1}{2}\|\rho_0-\rho_1\|_1\le \alpha$.
\item
Output: 
Accept if $\frac{1}{2}\|\rho_0-\rho_1\|_1\ge\beta$,
reject if $\frac{1}{2}\|\rho_0-\rho_1\|_1\le\alpha$.
\end{itemize}
Here,
$
\|X\|_1=\mbox{Tr}\sqrt{X^\dagger X}
$
is the trace norm.
It was shown in Ref.~\cite{WatrousQSZK} that if 
$0\le\alpha<\beta^2\le1$, the gap between $\alpha$ and $\beta$
can be amplified to $\alpha=2^{-r}$ and $\beta=1-2^{-r}$
for any polynomial $r$. Therefore, in this paper,
without loss of generality, we take
$\alpha=2^{-r+1}$ and $\beta=1-2^{-r+1}$
for any polynomial $r$.

The problem is a quantum version of the SZK-complete problem,
Statistical Difference~\cite{SV}.
The problem QSD$_{\alpha,\beta}$ and its
complement are QSZK-complete for any constants $\alpha$ and $\beta$ satisfying
$0<\alpha<\beta^2<1$~\cite{WatrousQSZK}.
In fact, as is shown in Ref.~\cite{WatrousQSZK}, the prover can
prove that two states $\rho_0$ and $\rho_1$ are far apart
in the following way.
\begin{itemize}
\item[1.]
The verifier uniformly randomly chooses $a\in\{0,1\}$,
and sends $\rho_a$ to the prover.
\item[2.]
The prover performs any measurement to distinguish $\rho_0$ and $\rho_1$,
and sends the result $a'\in\{0,1\}$ to the verifier.
\item[3.]
The verifier accepts if and only if $a=a'$.
\end{itemize}
Let $\{\Pi_0,\Pi_1\}$ be the POVM performed by the prover.
Then, the probability that the verifier accepts is
\begin{eqnarray*}
p_{acc}&=&\frac{1}{2}\mbox{Tr}(\Pi_0\rho_0)
+\frac{1}{2}\mbox{Tr}(\Pi_1\rho_1)\\
&=&\frac{1}{2}\mbox{Tr}(\Pi_0\rho_0)
+\frac{1}{2}\mbox{Tr}((I-\Pi_0)\rho_1)\\
&=&\frac{1}{2}+\frac{1}{2}\mbox{Tr}(\Pi_0\rho_0)
-\frac{1}{2}\mbox{Tr}(\Pi_0\rho_1)\\
&=&\frac{1}{2}+\frac{1}{2}\mbox{Tr}(\Pi_0(\rho_0-\rho_1))\\
&\le&\frac{1}{2}+\frac{1}{4}\|\rho_0-\rho_1\|_1.
\end{eqnarray*}
Therefore, for the YES case, by taking the optimal POVM,
\begin{eqnarray*}
p_{acc}&=&\frac{1}{2}+\frac{1}{4}\|\rho_0-\rho_1\|_1\\
&\ge&\frac{1}{2}+\frac{1}{2}(1-2^{-r+1})\\
&=&1-2^{-r},
\end{eqnarray*}
and for the NO case, for any POVM,
\begin{eqnarray*}
p_{acc}&\le&\frac{1}{2}+\frac{1}{2}2^{-r+1}\\
&=&\frac{1}{2}+2^{-r}.
\end{eqnarray*}

\begin{figure}[htbp]
\begin{center}
\includegraphics[width=0.5\textwidth]{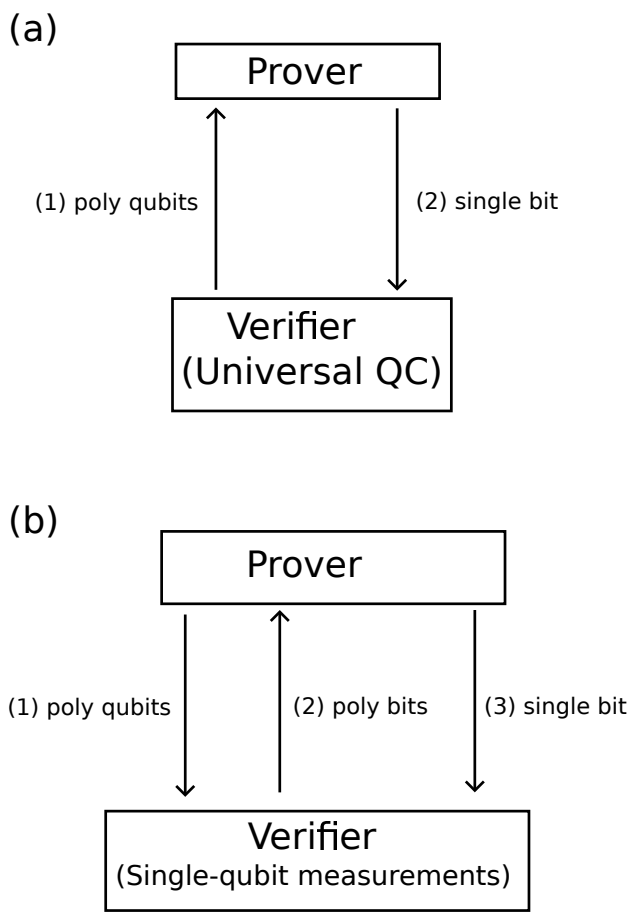}
\end{center}
\caption{
(a) The protocol of Ref.~\cite{WatrousQSZK}.
The verifier is quantum universal.
(b) Our protocol for QSD. The verifier does only
single-qubit measurements.
} 
\label{difference}
\end{figure}

The first result of the present paper is that QSD can be solved
with the verifier who can do only single-qubit measurements. 
The idea is that the honest prover generates
the graph state and sends a part of it to the verifier.
The verifier can remotely generates $\rho_0$ or $\rho_1$
in the prover's place by measuring his part.
The verifier can also check that his part is the correct graph state
by measuring stabilizer operators.
A trade-off is that, as is shown in Fig.~\ref{difference},
in the above protocol, one polynomial-size quantum message
from the verifier to the prover and one single-bit classical message
from the prover to the verifier are enough,
whereas in our protocol, one polynomial-size quantum
message from the prover to the verifier, one polynomial-size
classical message from the verifier to the prover,
and a single-bit classical message from the prover to the verifier
are necessary.

\subsection{QCD}
{\bf Definition}: Quantum Circuit Distinguishability 
(QCD$_{a,b}$)~\cite{Rosgen}.
\begin{itemize}
\item
Input: mixed-state quantum circuits, $Q_0$ and $Q_1$, both
of $n$-qubit input $m$-qubit output.
\item
Yes: $\|Q_0-Q_1\|_\diamond\ge a$.
\item
No: $\|Q_0-Q_1\|_\diamond\le b$.
\end{itemize}
Here, 
\begin{eqnarray*}
\|Q_0-Q_1\|_\diamond\equiv
\max_{X:\|X\|_1=1}\Big\|(Q_0\otimes I^{\otimes n})(X)
-(Q_1\otimes I^{\otimes n})(X)
\Big\|_1
\end{eqnarray*}
is the diamond norm.
It was shown in Ref.~\cite{Rosgen} that QCD$_{2-\delta,\delta}$ is
QIP-complete for any $\delta>0$.
In fact, the prover can proof that $Q_0$ and $Q_1$ are far apart
in the diamond norm as follows.
As is shown in Ref.~\cite{Rosgen}, there is a state $|\psi\rangle$
such that
\begin{eqnarray*}
\|Q_0-Q_1\|_\diamond=\Big\|
(Q_0\otimes I^{\otimes s})(|\psi\rangle\langle\psi|)
-(Q_1\otimes I^{\otimes s})(|\psi\rangle\langle\psi|)
\Big\|_1.
\end{eqnarray*}
For the YES case, the prover sends a part of
$|\psi\rangle$ to the verifier.
The verifier uniformly randomly chooses $i\in\{0,1\}$
and applies $Q_i$ on the part, and returns the state to the prover.
The prover now has $(Q_i\otimes I)(|\psi\rangle\langle\psi|)$,
and therefore he can learn $i$ by doing a measurement on the state
with the probability 
$\frac{1}{2}+\frac{1}{4}\|Q_0-Q_1\|_\diamond
\ge\frac{1}{2}+\frac{a}{4}$.
For the NO case, whatever state the prover sends to the verifier,
the acceptance probability is
less than
$\frac{1}{2}+\frac{1}{4}\|Q_0-Q_1\|_\diamond
\le\frac{1}{2}+\frac{b}{4}$.

\begin{figure}[htbp]
\begin{center}
\includegraphics[width=0.5\textwidth]{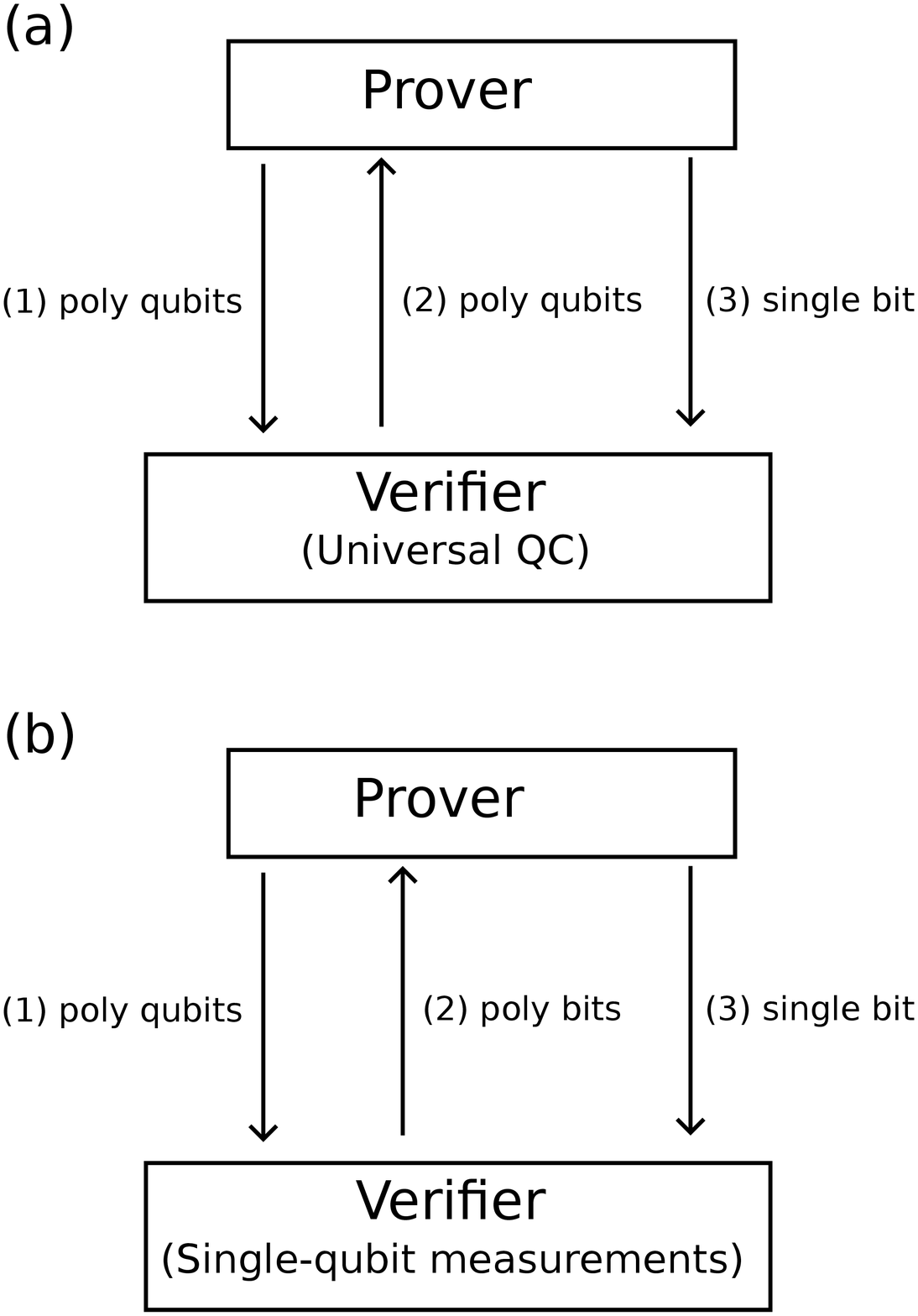}
\end{center}
\caption{
(a) The protocol of Ref.~\cite{Rosgen}.
The verifier is quantum universal.
(b) Our protocol for QCD.
The verifier does only single-qubit measurements.
} 
\label{difference2}
\end{figure}

Our second result is that QCD can be solved by the verifier
who can perform only single-qubit measurements.
As is shown in Fig.~\ref{difference2}, our protocol has an advantage that
the second quantum message from the verifier to the prover
can be replaced with the classical message, as well
as the fact that the verifier needs only single-qubit measurements.

Let us define the class QIP$_{\rm single}$ that is equivalent
to QIP except that the verifier can perform only single-qubit measurements.
Since quantum computing with measurements can be simulated by
a unitary quantum computing, it is obvious that
${\rm QIP}_{\rm single}\subseteq{\rm QIP}$.
On the other hand, our protocol that solves QCD is obviously in 
QIP$_{\rm single}$, and therefore our result means
${\rm QIP}\subseteq{\rm QIP}_{\rm single}$.
Hence, we have the result that
${\rm QIP}={\rm QIP}_{\rm single}$.
The result ${\rm QMA}_{\rm single}={\rm QMA}$
was shown in Ref.~\cite{MNS}, and
the result ${\rm QAM}_{\rm single}={\rm QAM}$
was shown in Ref.~\cite{QAMsingle}.
It was a remaining open problem whether 
${\rm QIP}_{\rm single}={\rm QIP}$.
The present paper solves it.

\section{Measurement-based quantum computing}
\label{Sec:MBQC}
For readers who are not familiar with measurement-based quantum
computing~\cite{MBQC}, we here explain basics of it.
Let us consider a graph $G=(V,E)$, 
where $|V|=N$.
The graph state $|G\rangle$ on $G$ is defined by
\begin{eqnarray*}
|G\rangle\equiv
\Big(\prod_{(i,j)\in E}CZ_{i,j}\Big)
|+\rangle^{\otimes N},
\end{eqnarray*}
where $|+\rangle\equiv(|0\rangle+|1\rangle)/\sqrt{2}$ and
$CZ_{i,j}\equiv|0\rangle\langle0|\otimes I+|1\rangle\langle1|\otimes Z$ 
is the CZ gate on the vertices $i$ and $j$.

According to the theory of measurement-based quantum computing~\cite{MBQC},
for any $m$-width $d$-depth quantum circuit $U$,
there exists a graph $G=(V,E)$ with $|V|=N=poly(m,d)$
and the graph state $|G\rangle$
on it such that if we measure each qubit 
in $V-V_o$, where $V_o$ is a certain
subset of $V$ with $|V_o|=m$, in certain bases adaptively,
then the state of $V_o$ after the measurements
is
\begin{eqnarray*}
B_{x,z}^mU|0^m\rangle
\end{eqnarray*}
with uniformly randomly chosen 
$x\equiv(x_1,...,x_m)\in\{0,1\}^m$ 
and $z\equiv(z_1,...,z_m)\in\{0,1\}^m$,
where 
\begin{eqnarray*}
B_{x,z}^m\equiv\bigotimes_{j=1}^m
X_j^{x_j}Z_j^{z_j}.
\end{eqnarray*}
This operator is called a byproduct operator,
and its effect is corrected, since $x$ and $z$ can be
calculated from measurement results.
Hence we finally obtain the desired state $U|0^m\rangle$.

The graph state $|G\rangle$ is stabilized by
\begin{eqnarray}
g_j\equiv X_j\bigotimes_{i\in S_j}Z_i,
\label{stabilizer}
\end{eqnarray}
for all $j\in V$, where $S_j$ is the set of
nearest-neighbour vertices of $j$th vertex. 
In other words,
\begin{eqnarray*}
g_j|G\rangle=|G\rangle
\end{eqnarray*}
for all $j\in V$. 

For $u\equiv(u_1,...,u_N)\in\{0,1\}^N$,
we define the state $|G_u\rangle$
by
\begin{eqnarray*}
g_j|G_u\rangle=(-1)^{u_j}|G_u\rangle
\end{eqnarray*}
for all $j\in V$.
(Therefore, $|G\rangle=|G_{0^N}\rangle$.)
The set $\{|G_u\rangle\}_u$ is an
orthonormal basis of the $N$-qubit Hilbert space.
In fact, if $u\neq u'$, there exists $j$ such that $u_j\neq u_j'$.
Then,
\begin{eqnarray*}
\langle G_{u'}|G_u\rangle&=&
\langle G_{u'}|g_jg_j|G_u\rangle\\
&=& (-1)^{u_j+u_j'}\langle G_{u'}|G_u\rangle\\
&=& -\langle G_{u'}|G_u\rangle,
\end{eqnarray*}
and therefore $\langle G_{u'}|G_u\rangle=0$.

\section{Stabilizer test}
\label{Sec:ST}
\begin{figure}[htbp]
\begin{center}
\includegraphics[width=0.45\textwidth]{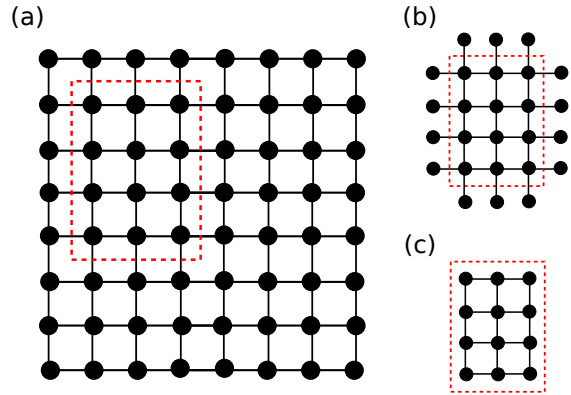}
\end{center}
\caption{
(a) The graph $G$. $V_1$ is the set of vertices in the
dotted red square, and  $V_2$ is the set of other vertices.
(b) The subgraph $G'$.
(c) The subgraph $G''$.
} 
\label{stabilizer_test}
\end{figure}

We now explain the stabilizer test. (See also Refs.~\cite{HM,MNS,QAMsingle}.)
Consider the graph $G=(V,E)$ of Fig.~\ref{stabilizer_test}.
(For simplicity, we here consider the square lattice,
but the result can be applied to any reasonable graph.)
As is shown in Fig.~\ref{stabilizer_test}, 
we define two subsets, $V_1$ and $V_2\equiv V-V_1$, of $V$,
where $|V_1|=N_1$ and $|V_2|=N_2$.
We also define a subset $V_{connect}$ of $V_2$ by
\begin{eqnarray*}
V_{connect}\equiv
\{j\in V_2|\exists i\in V_1~
\mbox{s.t.}~(i,j)\in E\}.
\end{eqnarray*}
In other words, $V_{connect}$ is the set of vertices in $V_2$ that are
connected to vertices in $V_1$.
We further define two subsets of $E$:
\begin{eqnarray*}
E_1&\equiv&\{(i,j)\in E|i\in V_1~\mbox{and}~j\in V_1\},\\
E_{connect}&\equiv&\{(i,j)\in E|i\in V_1~\mbox{and}~j\in V_2\}.
\end{eqnarray*}
Finally, we define two subgraphs
of $G$:
\begin{eqnarray*}
G'&\equiv&(V_1\cup V_{connect},E_1\cup E_{connect}),\\
G''&\equiv&(V_1,E_1).
\end{eqnarray*}

The stabilizer test is the following test:
\begin{itemize}
\item[1.]
Randomly generate an $N_1$-bit string 
$k\equiv(k_1,...,k_{N_1})\in\{0,1\}^{N_1}$.
\item[2.]
Measure the operator
\begin{eqnarray*}
s_k\equiv\prod_{j\in V_1}(g_j')^{k_j},
\end{eqnarray*}
where
$g_j'$ is the stabilizer operator, Eq.~(\ref{stabilizer}), of the graph state
$|G'\rangle$.
\item[3.]
If the result is $+1$ $(-1)$, the test passes (fails).
\end{itemize}
Let $|\Psi\rangle$ be a pure state on $V$.
If the probability $p_{pass}$ that $|\Psi\rangle$
passes the stabilizer test
satisfies $p_{pass}\ge1-\epsilon$, then
\begin{eqnarray}
\frac{1}{2}
\Big\||\Psi\rangle\langle\Psi|
-|\Psi'\rangle\langle\Psi'|
\Big\|_1\le\sqrt{4\epsilon-4\epsilon^2},
\label{st}
\end{eqnarray}
where 
\begin{eqnarray*}
|\Psi'\rangle\equiv
W(|G''\rangle\otimes|\xi\rangle_{V_2}).
\end{eqnarray*}
Here, $|\xi\rangle$ is a certain state on $V_2$
and
\begin{eqnarray*}
W\equiv\prod_{(i,j)\in E_{connect}}CZ_{i,j}.
\end{eqnarray*}

The proof is given as follows.
The probability $p_{test}$ that
the state $|\Psi\rangle$ on $V$
passes the stabilizer test is
\begin{eqnarray*}
p_{test}=\frac{1}{2^{N_1}}\sum_{k\in\{0,1\}^{N_1}}
\langle\Psi|\frac{I+s_k}{2}|\Psi\rangle.
\end{eqnarray*}
If we use the relation
\begin{eqnarray*}
\prod_{j\in V_1}\frac{I+g_j'}{2}
=\frac{1}{2^{N_1}}\sum_{k\in\{0,1\}^{N_1}}s_k,
\end{eqnarray*}
the condition $p_{test}\ge1-\epsilon$ means
\begin{eqnarray}
\langle\Psi|\prod_{j\in V_1}
\frac{I+g_j'}{2}|\Psi\rangle\ge1-2\epsilon.
\label{means}
\end{eqnarray}
Let $\{|\phi_t\rangle\}_t$ be an orthonormal basis
of $N_2$-qubit Hilbert space,
where $t\in\{0,1\}^{N_2}$.
Then, $\{W|G''_u\rangle\otimes|\phi_t\rangle\}_{u,t}$
is an orthonormal basis of the $N$-qubit Hilbert space,
and therefore,
$|\Psi\rangle$ can be written as
\begin{eqnarray*}
|\Psi\rangle=\sum_{u,t}
C_{u,t}W|G_u''\rangle\otimes|\phi_t\rangle,
\end{eqnarray*}
for certain coefficients $\{C_{u,t}\}_{u,t}$. 
Let us define
\begin{eqnarray*}
|\Psi'\rangle\equiv
W|G''\rangle\otimes\Big(
\frac{1}{\sqrt{R}}
\sum_t
C_{0^{N_1},t}|\phi_t\rangle\Big),
\end{eqnarray*}
where 
\begin{eqnarray*}
R\equiv\sum_t
|C_{0^{N_1},t}|^2\le1
\end{eqnarray*}
is the normalization constant.

Let $\{g_j''\}_j$ be the set of stabilizer operators of the graph state
$|G''\rangle$.
Then, it is easy to check
\begin{eqnarray*}
g_j' W=Wg_j''
\end{eqnarray*}
for all $j\in V_1$.
Therefore,
\begin{eqnarray*}
\Big(\prod_{j\in V_1}
\frac{I+g_j'}{2}\Big)|\Psi\rangle
&=&
W\prod_{j\in V_1}
\frac{I+g_j''}{2}
\Big(\sum_{u,t}C_{u,t}|G_u''\rangle\otimes|\phi_t\rangle\Big)\\
&=&
W\Big(\sum_tC_{0^{N_1},t}|G''\rangle\otimes|\phi_t\rangle\Big)\\
&=&\sqrt{R}|\Psi'\rangle.
\end{eqnarray*}
Hence Eq.~(\ref{means}) means
\begin{eqnarray*}
1-2\epsilon&\le&
\sqrt{R}\langle\Psi|\Psi'\rangle\\
&\le&\langle\Psi|\Psi'\rangle.
\end{eqnarray*}
Therefore,
\begin{eqnarray*}
\frac{1}{2}\Big\|
|\Psi\rangle\langle\Psi|
-|\Psi'\rangle\langle\Psi'|
\Big\|_1
&=&\sqrt{1-|\langle\Psi|\Psi'\rangle|^2}\\
&\le&\sqrt{1-(1-2\epsilon)^2}\\
&=&\sqrt{4\epsilon-4\epsilon^2}.
\end{eqnarray*}

\section{QSD}
\label{Sec:QSD}
\begin{figure}[htbp]
\begin{center}
\includegraphics[width=0.3\textwidth]{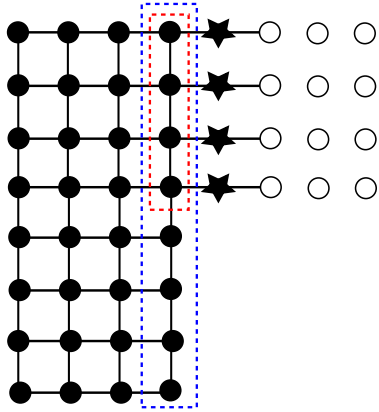}
\end{center}
\caption{
The graph $G$ for our protocol solving QSD.
} 
\label{fig2}
\end{figure}

In this section, we explain our protocol for QSD.
Let us consider the graph $G=(V,E)$ of Fig.~\ref{fig2}.
Our protocol runs as follows:
\begin{itemize}
\item[1.]
The prover generates a state $|\Psi\rangle$ on
$V$, 
and sends all black qubits
to the verifier.
If the prover is honest, 
$|\Psi\rangle\equiv|G\rangle$.
If the prover is malicious, $|\Psi\rangle$ can be any state.
\item[2.]
With probability $q$, which is specified later,
the verifier does the following.
\begin{itemize}
\item[2-a]
The verifier uniformly randomly chooses $a\in\{0,1\}$.
\item[2-b]
The verifier performs the measurement-based quantum computing
on the received qubits so that
the state of qubits in the blue dotted box becomes
$B_{x,z}^mQ_a|0^m\rangle$,
and the reduced state of the qubits in the red dotted box
becomes
$B_{x,z}^k\rho_aB_{x,z}^k$.
\item[2-c]
The verifier sends the prover $(x_1,...,x_k)$ and $(z_1,...,z_k)$.
\item[2-d]
The verifier measures qubits in the red dotted box
and the black star qubits
in the $X$ basis (in order to teleport the state 
to the white qubits that are connected to the star qubits),
and sends the $X$-basis measurement results to the prover.
\item[2-e]
The verifier receives the answer bit $a'\in\{0,1\}$ from the prover.
\item[2-f]
The verifier accepts if and only if $a=a'$.
\end{itemize}
We denote the acceptance probability
by $p_{comp}$. 
\item[3.]
With probability $1-q$, the verifier does the stabilizer test
by considering $V_1$ as the set of black circle qubits.
The verifier accepts if and only if the stabilizer test passes.
We denote the acceptance probability by $p_{test}$.
\end{itemize}

First, let us consider the YES case, i.e.,
$\frac{1}{2}\|\rho_0-\rho_1\|_1\ge1-2^{-r+1}$. In this case,
the prover is honest, and therefore $|\Psi\rangle=|G\rangle$,
which means $p_{test}=1$ if the verifier chooses the stabilizer test.
If the verifier chooses the computation,
after the all verifier's measurements, the state of the
white qubits that are connected to the star qubits becomes
$B_{x',z'}^k\rho_aB_{x',z'}^k$, where $x'$ and $z'$ can be
calculated from the all classical information from the verifier.
Therefore, the prover finally has $\rho_a$,
and the prover can learn $a$ by doing an appropriate POVM
with an error probability less than $2^{-r}$.
Hence the acceptance probability $p_{acc}$ of the protocol is
\begin{eqnarray*}
p_{acc}&=&qp_{comp}+(1-q)p_{test}\\
&\ge& q(1-2^{-r})+(1-q)\equiv\alpha.
\end{eqnarray*}

Second, let us consider the NO case,
namely, $\frac{1}{2}\|\rho_0-\rho_1\|_1\le 2^{-r+1}$.
If $p_{pass}<1-\epsilon$, where $\epsilon$ is a certain
parameter that will be specified later, there is no guarantee that
the prover generated the correct graph state.
Therefore,
$p_{comp}=1$ in the worst case:
\begin{eqnarray*}
p_{acc}&=&qp_{comp}+(1-q)p_{test}\\
&\le& q+(1-q)(1-\epsilon)\equiv\beta_1.
\end{eqnarray*}

If $p_{pass}\ge1-\epsilon$, on the other hand, 
$|\Psi\rangle$ is close to 
\begin{eqnarray*}
|\Psi'\rangle\equiv W(|G''\rangle\otimes|\xi\rangle)
\end{eqnarray*}
in the sense of Eq.~(\ref{st}), 
where $|\xi\rangle$ is a state on the star and white qubits,
and $W$ is the unitary operator that applies $CZ$ gates
on all edges that connect the qubits in the dotted red box
and star qubits.
For simplicity, let us assume that $|\Psi\rangle=|\Psi'\rangle$
for the moment. 
Then, after the step 2-c of the protocol, 
the state of qubits in the red dotted box, star qubits, white qubits,
and prover's classical memory 
is
\begin{eqnarray*}
W(B_{x,z}^k\rho_aB_{x,z}^k\otimes|\xi\rangle\langle\xi|)W^\dagger
\otimes|x,z\rangle\langle x,z|.
\end{eqnarray*}
However, since
\begin{eqnarray*}
&&
\frac{1}{2}\Big\|
W(B_{x,z}^k\rho_0B_{x,z}^k\otimes|\xi\rangle\langle\xi|)W^\dagger
\otimes|x,z\rangle\langle x,z|
-W(B_{x,z}^k\rho_1B_{x,z}^k\otimes|\xi\rangle\langle\xi|)W^\dagger
\otimes|x,z\rangle\langle x,z|
\Big\|_1\\
&=&\frac{1}{2}\|\rho_0-\rho_1\|_1\\
&\le&2^{-r+1},
\end{eqnarray*}
no POVM can distinguish
$\rho_0$ and $\rho_1$ with a probability
larger than $\frac{1}{2}+2^{-r}$.
Therefore, for any $|\Psi\rangle$ that satisfies
$p_{test}\ge1-\epsilon$, the acceptance probability is
\begin{eqnarray*}
p_{acc}&=&
qp_{comp}+(1-q)p_{test}\\
&\le& q\Big(\frac{1}{2}+2^{-r}+\sqrt{4\epsilon-4\epsilon^2}\Big)
+(1-q)\equiv\beta_2.
\end{eqnarray*}
If we define
\begin{eqnarray*}
\Delta_1(q)&\equiv&\alpha-\beta_1=-q2^{-r}+\epsilon(1-q),\\
\Delta_2(q)&\equiv&\alpha-\beta_2=\frac{q}{2}-q2^{-r+1}
-q\sqrt{4\epsilon-4\epsilon^2},
\end{eqnarray*}
then the optimal value $q^*$ of $q$, which satisfies
$\Delta_1(q^*)=\Delta_2(q^*)$, is
\begin{eqnarray*}
q^*\equiv\frac{\epsilon}{\epsilon+\frac{1}{2}-2^{-r}
-\sqrt{4\epsilon-4\epsilon^2}}
\end{eqnarray*}
and the gap for this $q^*$ is
\begin{eqnarray*}
\Delta_2(q^*)&=&
\frac{\epsilon(\frac{1}{2}-2^{-r+1}-\sqrt{4\epsilon-4\epsilon^2})}
{\epsilon+\frac{1}{2}-2^{-r}-\sqrt{4\epsilon-4\epsilon^2}}\\
&\ge&
\frac{\epsilon(\frac{1}{2}-\frac{1}{4}-\sqrt{4\epsilon})}
{\epsilon+\frac{1}{2}}\\
&=&
\frac{\frac{1}{2}-\frac{1}{4}-\frac{1}{5}}
{1+50}\\
&=&\frac{1}{1020}
\end{eqnarray*}
if $r\ge3$ and $\epsilon=\frac{1}{100}$.

\section{QCD}
\label{Sec:QCD}
\begin{figure}[htbp]
\begin{center}
\includegraphics[width=0.4\textwidth]{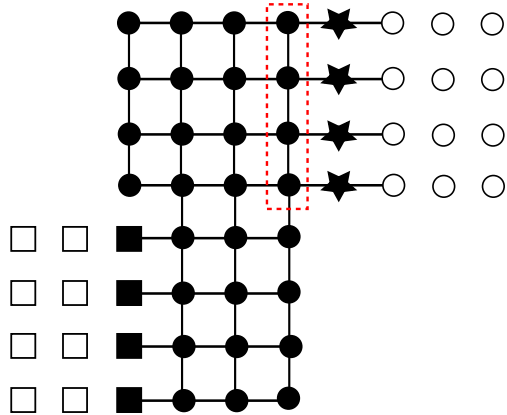}
\end{center}
\caption{
The graph $G$ for our protocol solving QCD.
} 
\label{QCD}
\end{figure}

In this section, we explain our protocol for QCD.
Let us consider the graph $G=(V,E)$ of Fig.~\ref{QCD}.
Our protocol runs as follows:
\begin{itemize}
\item[1.]
The prover generates a state $|\Psi\rangle$ on $V$
and sends all black qubits to the verifier.
If the prover is honest, 
\begin{eqnarray*}
|\Psi\rangle=W_1(|G_1\rangle\otimes|\psi\rangle),
\end{eqnarray*}
where $G_1$ is the subgraph of $G$ that is obtained
by removing all square vertices and all edges
that connect the black square vertices and black circle vertices,
$|\psi\rangle$ is the state of the square qubits (black
square qubits are those on which $Q_i$ should be acted),
and
$W_1$ is the unitary operator applying $CZ$ gates on
all edges that connect the black square qubits and
black circle qubits.
If the prover is malicious, $|\Psi\rangle$ can be any state.
\item[2.]
With probability $q$, which is specified later, the verifier
does the following.
\begin{itemize}
\item[2-a]
The verifier uniformly randomly chooses $i\in\{0,1\}$.
\item[2-b]
The verifier does the measurement-based quantum computation
so that the black circle qubits in the dotted red box
and white square qubits becomes 
\begin{eqnarray*}
(B_{x,z}^m\otimes I)
[
(Q_i\otimes I)(|\psi\rangle\langle\psi|)
]
(B_{x,z}^m\otimes I).
\end{eqnarray*}
\item[2-c]
The verifier sends $x$ and $z$ to the prover.
\item[2-d]
The verifier measures the black circle qubits in the red dotted box
and black star qubits in the $X$ basis, and sends the measurement
results to the prover.
\item[2-e]
The verifier receives $j\in\{0,1\}$ from the prover.
The verifier accepts if and only if $i=j$. We denote the
acceptance probability by $p_{comp}$.
\end{itemize}
\item[3.]
With probability $1-q$, the verifier does the stabilizer test
by considering $V_1$ as the set of black circle qubits.
The verifier accepts if and only if the test passes.
We denote the acceptance probability by $p_{test}$.
\end{itemize}

First, let us consider the YES case,
i.e., $\|Q_0-Q_1\|_\diamond\ge a$.
In this case, the prover is honest, and therefore, $p_{test}=1$
and 
\begin{eqnarray*}
p_{comp}&=&\frac{1}{2}+\frac{1}{4}\|Q_0-Q_1\|_\diamond\\
&\ge&\frac{1}{2}+\frac{a}{4}.
\end{eqnarray*}
Therefore,
\begin{eqnarray*}
p_{acc}&=&qp_{comp}+(1-q)p_{test}\\
&\ge&q\Big(\frac{1}{2}+\frac{a}{4}\Big)+(1-q)\equiv\alpha.
\end{eqnarray*}

Next let us consider the NO case, i.e., 
$\|Q_0-Q_1\|_\diamond\le b$.
If $p_{test}<1-\epsilon$,
\begin{eqnarray*}
p_{acc}
&=&qp_{comp}+(1-q)p_{test}\\
&\le& q+(1-q)(1-\epsilon)\equiv\beta_1.
\end{eqnarray*}

If $p_{test}\ge1-\epsilon$, on the other hand,
$|\Psi\rangle$ is close to 
\begin{eqnarray*}
|\Psi'\rangle=W(|G''\rangle\otimes |\xi\rangle)
\end{eqnarray*}
in the sense of Eq.~(\ref{st}).
Here, $G''$ is the graph whose vertices are black circle
qubits and whose edges are those connecting black circle
qubits.
The operator $W$ is the unitary operator applying $CZ$ gates
on all edges that connect black circle qubits and
the black star or black square qubits.
The state $|\xi\rangle$ is the state of the black star qubits,
black square qubits, and white qubits.
For the moment, let us assume that $|\Psi\rangle=|\Psi'\rangle$.
After the step 2-c, the state of white qubits, star qubits,
black circle qubits in the red dotted box, and prover's classical
memory is
\begin{eqnarray*}
[
W_2
(B_{x,z}^m\otimes I)
(Q_i\otimes I)
(|\xi\rangle\langle\xi|)
(B_{x,z}^m\otimes I)
W_2^\dagger
]
\otimes|x,z\rangle\langle x,z|,
\end{eqnarray*}
where $W_2$ is the unitary operator applying $CZ$ gates
on all edges that connects black circle qubits in the red dotted box
and star qubits.
However,
\begin{eqnarray*}
&&\Big\|
[W_2(B_{x,z}^m\otimes I)
(Q_0\otimes I)
(|\xi\rangle\langle\xi|)
(B_{x,z}^m\otimes I)W_2^\dagger
]\otimes|x,z\rangle\langle x,z|\\
&&-[W_2(B_{x,z}^m\otimes I)
(Q_1\otimes I)
(|\xi\rangle\langle\xi|)
(B_{x,z}^m\otimes I)W_2^\dagger
]\otimes|x,z\rangle\langle x,z|
\Big\|_1\\
&=&
\Big\|
(Q_0\otimes I)(|\xi\rangle\langle\xi|)
-(Q_1\otimes I)(|\xi\rangle\langle\xi|)
\Big\|_1\\
&\le&
\|
Q_0-Q_1
\|_\diamond,
\end{eqnarray*}
and therefore, $p_{comp}\le \frac{1}{2}+\frac{b}{4}$.
Hence for any $|\Psi\rangle$ such that $p_{test}\ge1-\epsilon$,
the acceptance probability is
\begin{eqnarray*}
p_{acc}
&=&qp_{comp}+(1-q)p_{test}\\
&\le&q\Big(\frac{1}{2}+\frac{b}{4}+\sqrt{4\epsilon-4\epsilon^2}\Big)
+(1-q)\equiv\beta_2.
\end{eqnarray*}
If we define
\begin{eqnarray*}
\Delta_1(q)&\equiv&\alpha-\beta_1
=-\frac{q}{2}+\frac{qa}{4}+\epsilon(1-q),\\
\Delta_2(q)&\equiv&\alpha-\beta_2=
\frac{q(a-b)}{4}-q\sqrt{4\epsilon-4\epsilon^2},
\end{eqnarray*}
the optimal value $q^*$ of $q$ is
\begin{eqnarray*}
q^*\equiv\frac{\epsilon}{\frac{1}{2}+\epsilon
-\sqrt{4\epsilon-4\epsilon^2}-\frac{b}{4}},
\end{eqnarray*}
and the gap is
\begin{eqnarray*}
\Delta_2(q^*)
&=&
\frac{\epsilon(\frac{a-b}{4}-\sqrt{4\epsilon-4\epsilon^2})}
{\frac{1}{2}+\epsilon
-\sqrt{4\epsilon-4\epsilon^2}-\frac{b}{4}}\\
&\ge&
\frac{\epsilon(\frac{a-b}{4}-2\sqrt{\epsilon})}
{\frac{1}{2}+\epsilon}\\
&=&\frac{1}{1020}
\end{eqnarray*}
if we take $\epsilon=\frac{1}{100}$,
$a=1.5$, and $b=0.5$.
Note that the error can be reduced by running the protocol in parallel,
and using the Markov inequality argument~\cite{JUW}.

\acknowledgements
The author acknowledges Harumichi Nishimura,
Hirotada Kobayashi, and Adam Bouland
for discussion, and 
Grant-in-Aid for Scientific Research on Innovative Areas
No.15H00850 of MEXT Japan, and the Grant-in-Aid
for Young Scientists (B) No.26730003 of JSPS
for the support.

\end{document}